# The Earliest Candidates of Auroral Observations in Assyrian Astrological Reports: Insights on Solar Activity around 660 BCE

Hisashi Hayakawa (1-2)*, Yasuyuki Mitsuma (3)**, Yusuke Ebihara (4), and Fusa Miyake (5)

(1) Graduate School of Letters, Osaka University, 5600043, Toyonaka, Japan (JSPS Research Fellow)
(2) Science and Technology Facilities Council, RAL Space, Rutherford Appleton Laboratory, Harwell Campus, Didcot, OX11 0QX, UK
(3) Faculty of Humanities and Social Sciences, University of Tsukuba, Tsukuba, 3058571 Japan
(4) Research Institute for Sustainable Humanosphere, Kyoto University, Uji, 6100011, Japan
(5) Institute for Space-Earth Environmental Research, Nagoya University, 4648601, Nagoya, Japan

* hayakawa@kwasan.kyoto-u.ac.jp; hisashi.hayakawa@stfc.ac.uk
** licornesoleil@gmail.com

**Abstract**

Auroral records found in historical archives and cosmogenic isotopes found in natural archives have served as sound proxies of coronal mass ejections (CMEs) and solar energetic particles (SEPs), respectively, for dates prior to the onset of telescopic sunspot observations in 1610. These space weather events constitute a significant threat to a modern civilization, because of its increasing dependency on an electronic infrastructure. Recent studies have identified multiple extreme space weather events derived from solar energetic particles (SEPs) in natural archives, such as the event in 660 BCE. While the level of solar activity around 660 BCE is of great interest, this had not been within the coverage of the hitherto-known datable auroral records in historical documents that extend back to the 6th century BCE. Therefore, we have examined Assyrian astrological reports in the 8th and 7th centuries BCE, identified three observational reports of candidate aurorae, and dated these reports to approximately 680 BCE – 650 BCE. The Assyrian cuneiform tablets let us extend the history of auroral records and solar activity by a century. These cuneiform reports are considered to be the earliest datable records of candidate aurorae and they support the concept of enhanced solar activity suggested by the cosmogenic isotopes from natural archives.

## 1. Introduction

The Sun causes infrequent but high-impact space weather hazards. These space weather events represent a significant threat to modern society, as it becomes increasingly dependent on a technology-based infrastructure and is therefore very vulnerable to such hazards (Daglis, 2001; Riley *et al.*, 2018; Dyer *et al.*, 2018). These space weather hazards are caused by coronal mass ejections (CMEs) and solar energetic particles (SEPs) from our Sun (Daglis, 2001; Vaquero and Vázquez, 2009; Riley *et al.*, 2018; Dyer *et al.*, 2018), although their intensity correlation is at best moderate (Gopalswamy *et al.*, 2012).

Recent analyses of $^{14}C$ in tree rings have revealed anomalous enhancements in 774/775 CE and in 992/993 CE or 993/994 CE (Miyake *et al.*, 2012, 2013; Usoskin *et al.*, 2013; Büntgen *et al.*, 2018) and these enhancements have been associated with anomalous SEP events, on the basis of simultaneous enhancements of $^{14}C$ in other tree rings and enhancements of $^{10}Be$ and $^{36}Cl$ in polar ice cores (Mekhaldi *et al.*, 2015; Büntgen *et al.*, 2018). These events occurred far before the onset of instrumental observations, well outside the more modern range of wide observational coverage (see Vaquero and Vázquez, 2009; Clette *et al.*, 2014). Therefore, in order to infer the general trend of solar activity and the occurrence of CMEs, candidate auroral records have been sought in historical documents around these events (*e.g.*, Usoskin *et al.*, 2013).

Further analyses in the cosmogenic isotope data have identified a third extreme SEP event in 660 BCE (Park *et al.*, 2017; O'Hare *et al.*, 2019). This event occurred slightly before the established coverage of datable records of candidate aurorae back to the 6th century BCE (Stephenson *et al.*, 2004; Silverman, 2006; Hayakawa *et al.*, 2016; see Appendix 1).

However, the Babylonians and Assyrians had started astrological observations[1] at the latest in the 8th century BCE (Steele, 2012). Already in the 7th century BCE, Assyrian kings had collected and received astrological reports from professional astrologers, to interpret the ominous meaning of observed celestial events (Hunger, 1992). Therefore, in this contribution, we survey candidate auroral records in Assyrian astrological reports and compare them with the available scientific data, in order to extend our knowledge on variable solar activity and the occurrence of CMEs back into the 7th century BCE.

[1] Here, we use the terms of astronomy and astrology in the Assyrian epoch, based on their purpose: astronomy for calendar-setting and prediction of celestial phenomena and astrology for divinatory prediction of terrestrial events from immediate celestial phenomena.

## 2. Materials and Methods

Most of the extant reports were written during the 8th and 7th centuries BCE and were collected in the royal archives of Nineveh, the last capital of the Assyrian Empire after the reign of Sennacherib (r. 705 BCE–681 BCE). These reports were comprehensively collected and edited in Hunger (1992). Assyrian kings employed scholars specialised in divination. These scholars sent reports on celestial phenomena to their employer-kings. These reports describe the astrological interpretations of these phenomena. The interpretations are often accompanied by technical phrases in the *Enūma Anu Enlil* (EAE) and other omen collections. Most of the reports were written on palm-size rectangular tablets, and the texts are inscribed along the longer sides of the tablets (Hunger, 1992).

Hunger (1992) explains that reports of such ominous celestial events may involve the following six elements.

I Quotations of given omens from EAE
II Explanations of those omens
III Statements of celestial observations, occasionally including predictions
IV Letter-like messages
V Sender's name
VI Date

The first element is seen in almost all extant reports. This element is involved in almost all the astrological reports and cited from Enuma Anu Enlil, on the basis of what they observed. According to Hunger (1992, p. xvi), "In principle, no other information is required, since the protasis of a celestial omen always implies an observation. There would seem to be no need to explicitly repeat what was observed".

Occasionally, actual observations are recorded, and the reported ones are often different from the quoted omen protases (sentences with cited omens). Especially when the real celestial events seemed not to repeat the protases clearly, explanations and statements of observations would have been necessary. On the contrary, observational statements were not necessary when protases of celestial omens implied observed events (Hunger, 1992). In the report no. 288 edited by Hunger (1992), Nabû-iqiša reports the observation of conjunction of Jupiter and Mars only by quoting some omen protases concerning this kind of conjunction and calls for a ritual against this bad omen. Therefore, unless otherwise endorsed, we surmise that quoted protases indicate actual celestial events with a fair degree of precision.

Although the datable astrological reports range chronologically from 709 BCE to 649 BCE (Brown, 2000), we have only one from the 8th century BCE (no. 501) (Hunger, 1992). The earliest known Babylonian astronomical diary is dated in the same period (≈ 652 BCE) (Sachs, 1988; Hayakawa *et al.*, 2016). We found names of 14 Assyrian senders and 30 Babylonian scholars in the reports (Hunger, 1992). Note that Babylonia was under the rule of the Assyrian Empire in this epoch (until 626 BCE).

Scholars had made regular observations and reported their observational results to the Assyrian kings. Two reports (Hunger, 1992, nos. 7 and 79) clearly show this kind of activity by the scholars. In the first report, Issār-šumu-ēreš reported the observations for Mars twice or thrice in one day. It had set and could not be seen. Issār-šumu-ēreš anticipated the king's inquiry as to its ominous meaning and wrote his answer to negate the existence of such a sign in the report. In the second report, Nabû-aḫḫē-erība reported a continuous observation for the Moon on the 29th day of a certain month and probably on the next day. These accounts show that the astrological reports are based on the regular observations, as stated by Hunger (1992).

Scholars sent their reports from Assyria, the area around Nineveh (N36°21′, E43°09′), and from cities in Babylonia such as Babylon (N32°33′, E44°26′), Borsippa (N32°24′, E44°20′), Cutha (N32°46′, E44°20′), Dilbat (N32°10′, E44°28′), Ur (N30°58′, E46°06′), and Uruk (N31°20′, E45°38′) (Brown, 2000).

Scholars quoted the protases and stated the observations about unusual astronomical phenomena such as comets (*ṣallummû*), meteors (*kakkabu rabû*), and halos (*tarbāṣu*) (Hunger, 1992). These reports show the Assyrian scholars' rich Akkadian technical terms and deep knowledge of various celestial events.

Unfortunately, only a small fraction of extant reports were dated explicitly. However, as these reports were drafted at the time of occurrence of given celestial events, these events must have occurred during the tenure of the astrologers mentioned in the reports. Therefore, we can constrain their date ranges based on when the relevant astrologers were actively working under the Assyrian court.

## 3. Results

In order to survey candidate auroral reports around the 7th century BCE, we have consulted the astrological reports in Hunger's critical edition (Hunger, 1992) and noted all mentions of reddish luminous phenomena in the sky (*c.f.*, Stephenson *et al.*, 2004; Hayakawa *et al.*, 2016). Among the 389 astrological reports surveyed, we have found three containing auroral candidates: 'red glow' (R1 = Rm211), 'red cloud' (R2 = K748), and 'red sky' (R3 = 80-7-19,19) (Appendix 2). These three auroral candidates are significantly rare within the surviving astrological reports (≈ 0.8%), whose main focus is lunar and planetary phenomena (Hunger, 1992). These events are therefore considered sufficiently notable for the Assyrians and Babylonians to include them in astrological discussions. Table 1 provides their estimated date ranges, references, and tablet names in Table 1. After identifying these reports, we have examined the available original cuneiform tablets in the British Museum and made copies (Figure 1), transliterated the original texts, and translated them into English (see Appendix 2).

Table 1: Summary of tablets with candidate auroral records, giving their tablet shelf mark, estimated date range, name of observer, observation site, and figure number of our trace copy (Appendix 2).

| ID | Tablet | Date Range | Observer | Site | Figure |
|---|---|---|---|---|---|
| R1 | Rm211 | 679 BCE – 655 BCE | Issār-šumu-ēreš | Nineveh | 1(a) |
| R2 | K748 | 677 BCE – 666 BCE | Nabû-aḫḫē-erība | Nineveh | 1(b) |
| R3 | 80-7-19,19 | 679 BCE – 670 BCE | Zākiru | Babylon | N/A |

## 4. Dates and Descriptions of These Astrological Reports

The tablet R1 (tablet Rm211) involves the technical term, *akukūtu* 'red glow'. This term is also found in the known candidate auroral report for 12/13 March 567 BCE (Stephenson *et al.*, 2004; Hayakawa *et al.*, 2016; see Appendix 1). This astrological report describes a blaze of this red glow in the month of Sivan (III) along with its associated omen interpretation. There are cases of a red glow blazing at the zenith, during motion from south to north, being associated with a south wind. This is consistent with the auroral report at Barnstaple in 1872, when the aurora extended from the south to the zenith (Hall, 1872; see Appendix 3). Hunger (1992) has identified the tablet R1 (tablet Rm211) as one of the astrological reports of the scholar Issār-šumu-ēreš to the Assyrian kings Esarhaddon and Assurbanipal. We estimate the R1 date range as falling between 679 BCE and 655 BCE, according to the dates when Issār-šumu-ēreš was working in the Assyrian court (Hunger, 1992; Parpola, 1993; Pearce, 2000).

The tablet R2 (tablet K748) contains the term 'red cloud'. Aurorae were certainly described using the term "red cloud" even in 19th century, as in the case of the 1859 August storm (Loomis, 1860; see Appendix 3). The tablet R2 is one of the astrological reports from Nabû-aḫḫē-erība to the Assyrian kings Esarhaddon and Assurbanipal. As Nabû-aḫḫē-erība worked under the Assyrian court between 677 BCE and 666 BCE (Hunger, 1992; Parpola, 1993), the date range of the tablet K748 is estimated accordingly.

The tablet R3 (tablet 80-7-19,19) reports 'red covers the sky'. It is known that auroral displays were reported to have covered the entire sky during extreme space weather events (*e.g.*, Loomis, 1860; Hall, 1872; see Appendix 3). The tablet R3 is one of astrological reports from Zākiru, who was probably active in Babylon between *c.a.* 679 BCE and 670 BCE (Hunger, 1992; Groß, 2011). Therefore, its date range is defined accordingly.

## 5. Scales of Associated Magnetic Storms

Zākiru was a Babylonian scholar, and Issār-šumu-ēreš and Nabû-aḫḫē-erība were Assyrian scholars active at Nineveh, as summarized in Table 1. Although not very frequent, aurorae had certainly been visible in the Middle East during extreme magnetic storms. In particular, during the extreme storms in October 1870, aurorae were reported at Cairo and Baghdad (Silverman, 2006; Vaquero *et al.*, 2008); during the extreme storms in February 1872, aurorae were reported at Alexandria and Cairo, respectively (Silverman, 2008).

Moreover, the Middle East was closer to the north geomagnetic pole in the Assyrian epoch. While the geomagnetic north pole is situated in the region of North America today, it was situated in the region of Eurasia in the mid- to early-7th century BCE due to the secular variation of the geomagnetic field (Korte and Constable, 2011). Figure 2 shows the secular variation of magnetic latitudes (MLATs) for Babylonian and Assyrian cities represented by Nineveh (N36°21′, E043°09′), Babylon (N32°33′, E044°26′), and Ur (N30°58′, E046°06′). Their MLATs during the Assyrian and Babylonian period were ≈ 8° further poleward compared with their MLATs in the 1870s. Therefore, ancient aurorae would have been visible much more frequently in Mesopotamia during large magnetic storms than in the 1870s, when aurorae were observed in the Middle East (Silverman, 2006, 2008; Vaquero *et al.*, 2008).

Indeed, the MLATs of Assyroian and Babylonian cities in the 7th century BCE ranged from 35° to 41°, which is further poleward (magnetically) than Hokkaido (Japan) at recent times, where reddish aurorae are often observed during small and moderate magnetic storms (Shiokawa *et al.*, 2005). These recent Japanese auroral observations at mid to low magnetic latitudes usually indicate auroral luminosity near the horizon in the poleward direction. However, overhead aurorae in Hokkaido have been reported for great magnetic storms like the 1909 one (Dst ≈ −595 nT) (Hayakawa *et al.*, 2019). Therefore, the occurrence of aurorae over Assyrian and Babylonian cities in the interval 679 BCE – 655 BCE let us reasonably suppose that large interplanetary coronal mass ejections and subsequent strong magnetic storms occurred in this same interval.

These astrological reports describe reddish luminous phenomena in the sky: 'red glow', 'red cloud', and 'red sky'. The reddish colour is quite typical of low-latitude aurorae and stable auroral red (SAR) arcs with $O_I$ emission (630.0 nm). The former one, reddish aurora, is caused by low-energy electrons (≤ 100 eV) and is frequently accompanied by greenish emissions (557.7 nm) (Rees and Roble, 1975). The latter one, the SAR arc, appears during the main phase and recovery phase of magnetic storms, equatorward of the auroral oval, and extends further in the longitudinal directions for a few hours (Kozyra *et al.*, 1997) or even longer (Craven *et al.*, 1982). They are caused by downward heat flux (≈ 3000 K) or precipitation of low-energy particles (Kozyra *et al.*, 1997). Although they are generally not bright, some bright SAR arcs have been reported (≈ 13 kilo Rayleigh) during large magnetic storms (Baumgardner *et al.*, 2007). The SAR arcs can be distinguished from the reddish aurora by comparing with greenish emissions (555.7 nm) (Miyaoka *et al.*, 1990; Shiokawa *et al.*, 2005). However, it is, in general, difficult to distinguish them without spectroscopic instruments.

## 6. Chronology in Longer-Term Solar Variability

Figure 3 shows a comparison of $^{14}C$ concentrations in tree rings (Park *et al.*, 2017; Reimer *et al.*, 2013) and the date ranges of the three candidate auroral records: 679 BCE – 655 BCE (R1), 677 BCE – 666 BCE (R2), and 679 BCE – 670 BCE (R3). Although an extremely large SEP event causes a detectable $^{14}C$ increase, as seen in the 774/775 CE event, a normal $^{14}C$ variation is caused by a modulation of galactic cosmic rays, due to changes in the interplanetary magnetic field (Beer *et al.*, 2012). The large $^{14}C$ excursion from 810 BCE to 720 BCE corresponds a period of extremely low solar activity, called a grand solar minimum (Usoskin *et al.*, 2007).

After this grand solar minimum, $^{14}C$ concentrations decrease as solar activity recovers, and then a rapid $^{14}C$ increase is observed in ≈ 660 BCE. From multi-proxy analyses of $^{14}C$ in tree rings and $^{10}Be$ and $^{36}Cl$ in ice cores, the origin of this rapid $^{14}C$ increase is considered to be one or more extreme SEP event(s) that occurred during the period of increasing solar activity (Usoskin *et al.*, 2013; Mekhaldi *et al.*, 2015). Two of these records (R2 and R3) fall in this period and hence may be associated with this enhanced solar activity. R1 has a rather broad time range (679 BCE – 655 BCE) and covers both the enhanced solar activity during 680 BCE – 670 BCE and the anomalous enhancements of cosmogenic isotopes around 660 BCE. Although R1's exact date is an open question, these reports seem to provide snapshots of enhanced solar activity during 680 BCE – 670 BCE and possibly the period around the 660 BCE event (Park *et al.*, 2017; O'Hare *et al.*, 2019).

## 7. Conclusion

In this study, we have surveyed the Assyrian astrological reports for the 7th century BCE. We have identified three candidate auroral observations in the Assyrian astrological reports and identified their probable date ranges: 679 BCE – 655 BCE (R1: Rm211), 677 BCE – 666 BCE (R2: K748), and 679 BCE – 670BCE (R3: 80-7-19,19). Although we cannot further refine the date ranges at this stage, these Assyrian reports provide snapshots of solar activity in the early 7th century BCE around extreme SEP events (Park *et al.*, 2017; O'Hare *et al.*, 2019) and allow us to trace the history of solar activity back a century earlier than the earliest existing datable auroral reports in the 6th century BCE (Stephenson *et al.*, 2004; Silverman, 2006; Vaquero and Vázquez, 2009; Hayakawa *et al.*, 2016).

## Acknowledgement

This work was supported by Grant-in-Aids from the Ministry of Education, Culture, Sports, Science and Technology of Japan, Grant Number JP18H01254 (PI: H. Isobe), JP18K00987 (PI: Y. Mitsuma), JP15H05816 (PI: S. Yoden), JP15H03732 (PI: Y. Ebihara), and JP15H05815 (PI: Y. Miyoshi), and a Grant-in-Aid for JSPS Research Fellows JP17J06954 (PI: H. Hayakawa). We thank the Trustees of the British Museum and Staatlich Museen Berlin for kindly letting us research these original cuneiform tablets. We also thank David M. Willis for his helpful comments and suggestions on our manuscript.

## Author Contributions

HH prepared the manuscript and coordinated the discussions, YM analysed the cuneiform tablets, YE detailed the auroral science, and FM analysed the contemporary cosmogenic isotopes. All the authors read this manuscript and contributed to the discussions.

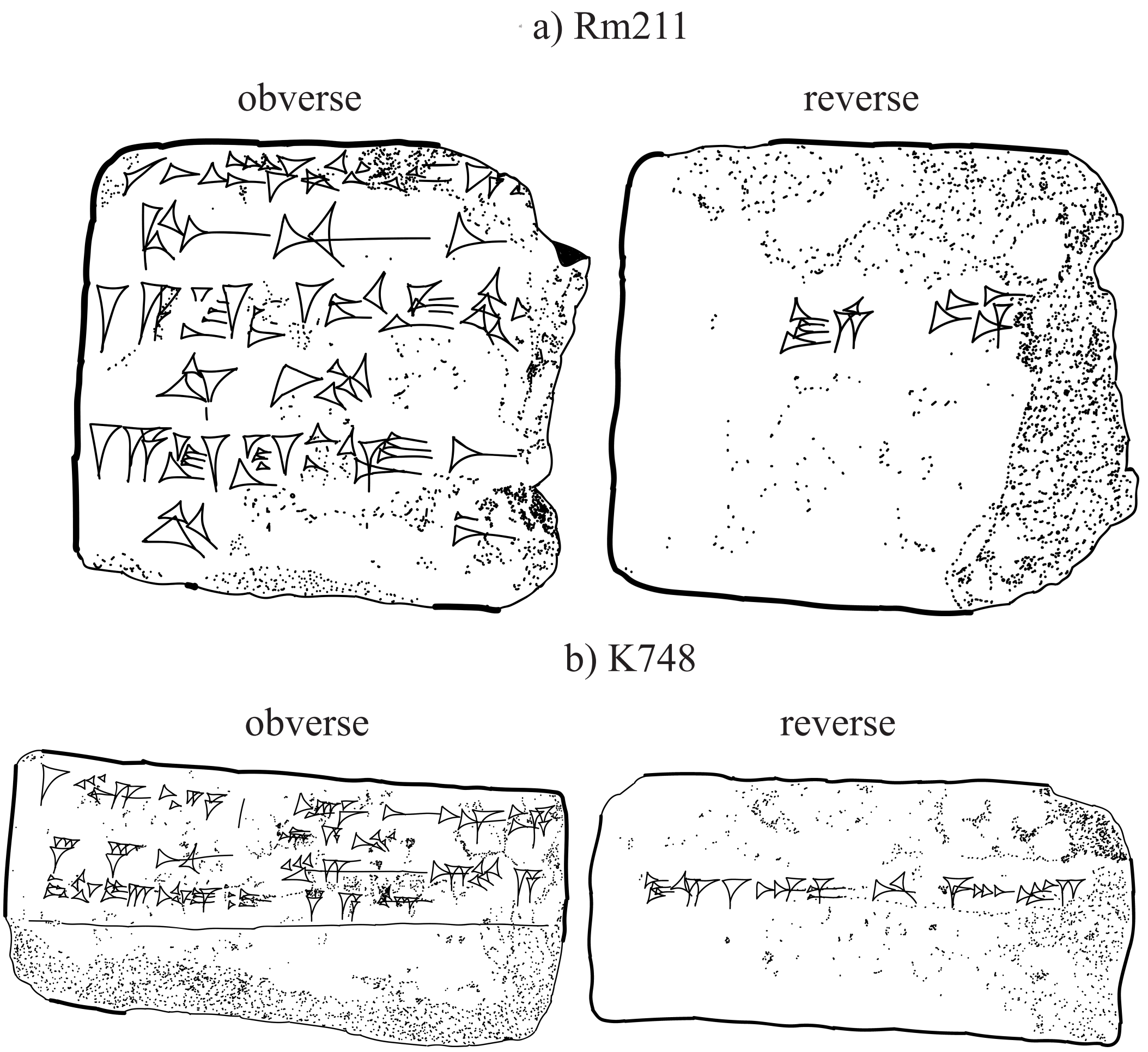

Figure 1: (above) R1 = Rm211 (Preserved in the British Museum, RMA 275; Hunger

1992, no. 35). Copies of 1 – 6 and r.1 (Y. Mitsuma's tracing of the photographs © H. Hayakawa, taken with courtesy of the Trustees of the British Museum); (below) R2 = K748 (Preserved in the British Museum, RMA 248; Hunger 1992, no. 78). Copy of 1 – 2 and r. 1 (Y. Mitsuma's tracing of the photograph © H. Hayakawa, taken courtesy of the Trustees of the British Museum)

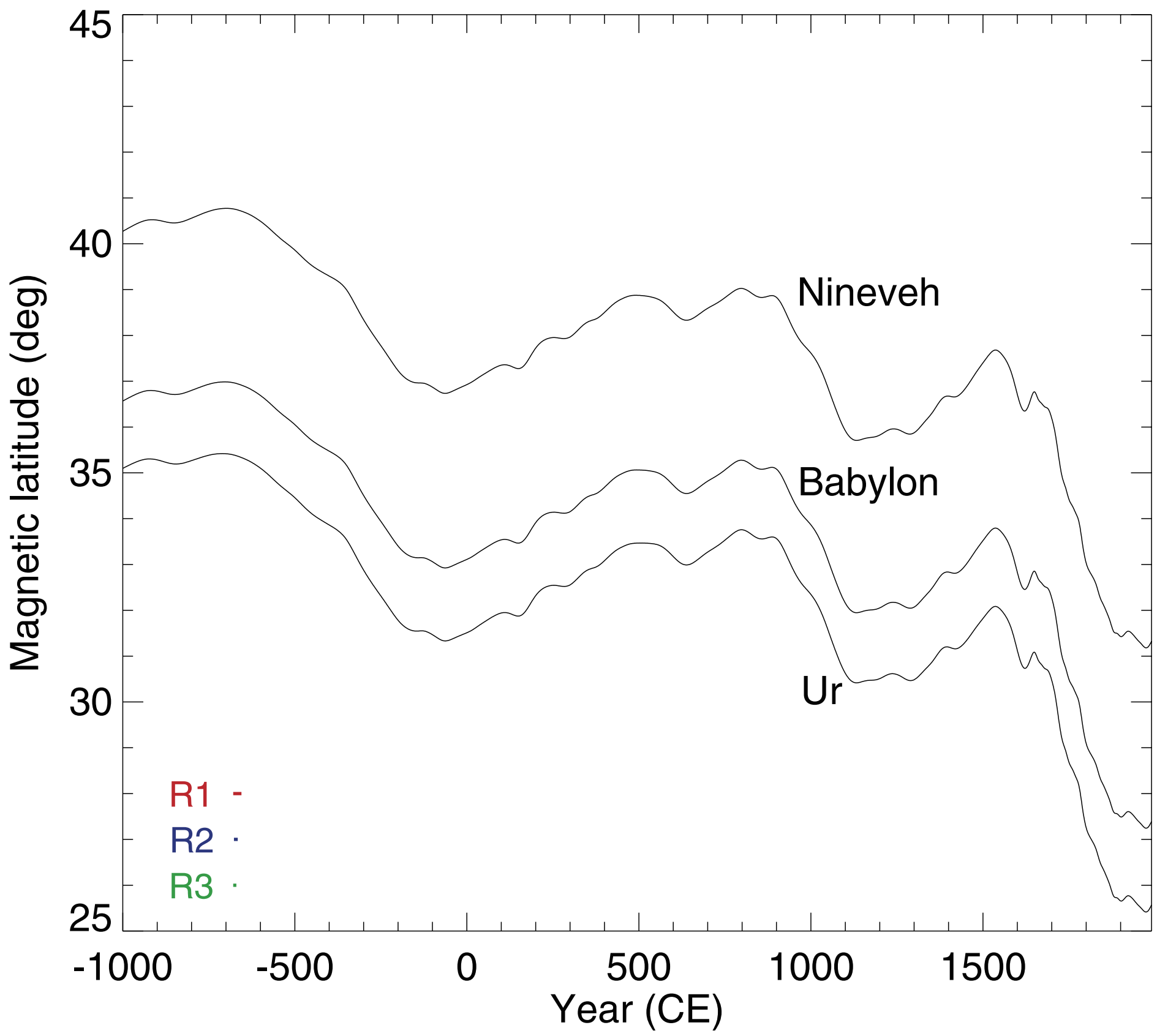

Figure 2: Secular variations of the MLATs of Babylonian and Assyrian cities, represented by Nineveh (N36°21′, E43°09′), Babylon (N32°33′, E44°26′), and Ur (N30°58′, E46°06′), as computed by the angular distance of the contemporary magnetic north pole with a dipole assumption in CALS3k4b (Korte and Constable, 2011).

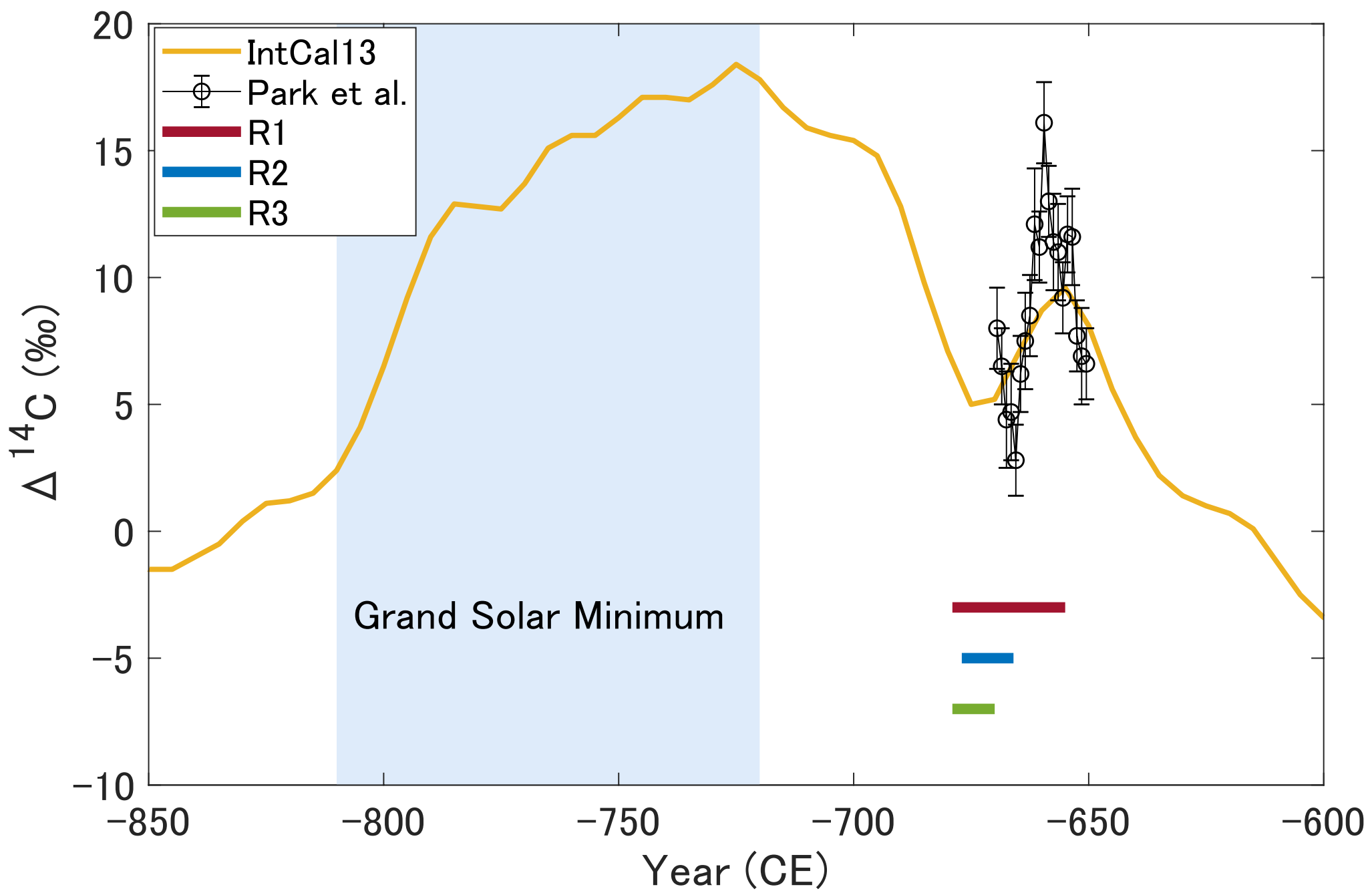

Figure 3: Carbon-14 concentrations for the period from 850 BCE to 600 BCE. Time resolutions are 5-year for the IntCal data (Reimer *et al.*, 2013) and 1-year for the data presented in Park *et al.* (2017), respectively. The period from 810 BCE to 720 BCE represents a grand solar minimum (Usoskin *et al.*, 2007). The timeframes of the three candidate auroral records are plotted together.

## Appendices

### Appendix 1: Existing Earliest Auroral Records

The earliest datable auroral record has been considered to be the Babylonian record on 12/13 March 567 BCE (VAT 4956; Stephenson *et al.*, 2004; Hayakawa *et al.*, 2016; Supplementary Figure S1). An astronomer-astrologer (*ṭupšar Enūma Anu Enlil*) reported this event as (see also Stephenson *et al.*, 2004; Hayakawa et al., 2016):

> *Transliteration*: GE$_6$ 29 *a-ku*$_6$*-ku*$_6$*-{ku*$_6$*}-tu*$_4$ *ina* ŠÚ KUR 2 DA[NNA ....]
> *Translation*: Night of the 29th, red glow flared up in the west; 2 double[-hours ....]

Its cuneiform text is shown in Figure S1. The MLAT of Babylon was computed to be 36.5° at that time (Korte and Constable, 2011).

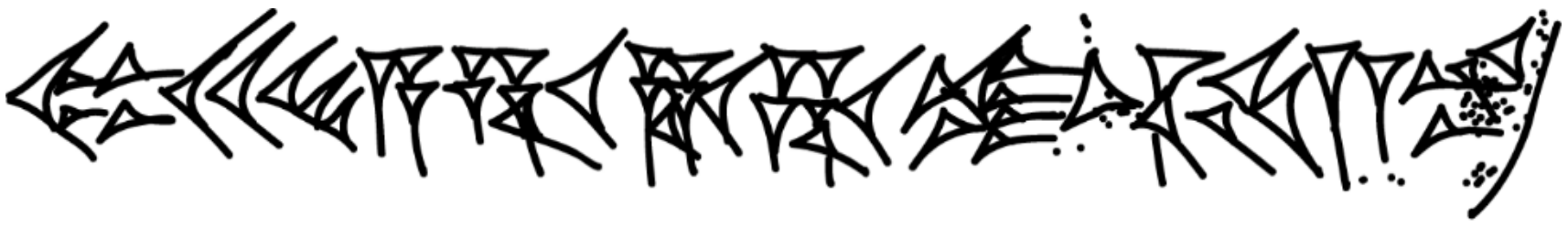

Supplementary Figure S1: VAT 4956, with the earliest datable auroral record in 12/13 March 567 BCE. (Tracing of the photographs © Hisashi Hayakawa, taken courtesy of the Staatliche Museen, Berlin).

Alternatively, some researchers have associated Ezekiel's Vision (Supplementary Figure S2; Kittel, 1906, p. 744) with an auroral display, as described in Ezekiel (1:4, 13-14) (*e.g.*, Silverman, 1998, 2006; Siscoe *et al.*, 2002). Silverman (2006) dates this event as *c.a.* 593–571 BCE and assumes its observational site to be Nippur (N32°08′, E045°14′), although this interpretation is still controversial (Stephenson *et al.*, 2004). Nevertheless, the analysis of Ezekiel (1:1-2) permits identification of the observational date. Ezekiel saw this vision on the fifth day of the fourth month in the fifth year of King Jehoiachin's captivity. Jehoiachin was captured by King Nebuchadnezzar II in his eighth year (2 Kings 24:12). Nebuchadnezzar's eighth year is considered to be 597/596 BCE on the basis of the accession-year system, but 598/597 BCE if his "first year" corresponds to the accession year (Finegan, 1950, p. 64). Therefore, the vision was probably seen on 12/13 July 594 BCE, or 30/31 July 593 BCE, according to the existing date conversion table (Parker and Dubberstein 1956, p. 28). The MLAT of Nippur is computed to have been ≈ 36.2° at that time with CALS3k4b.

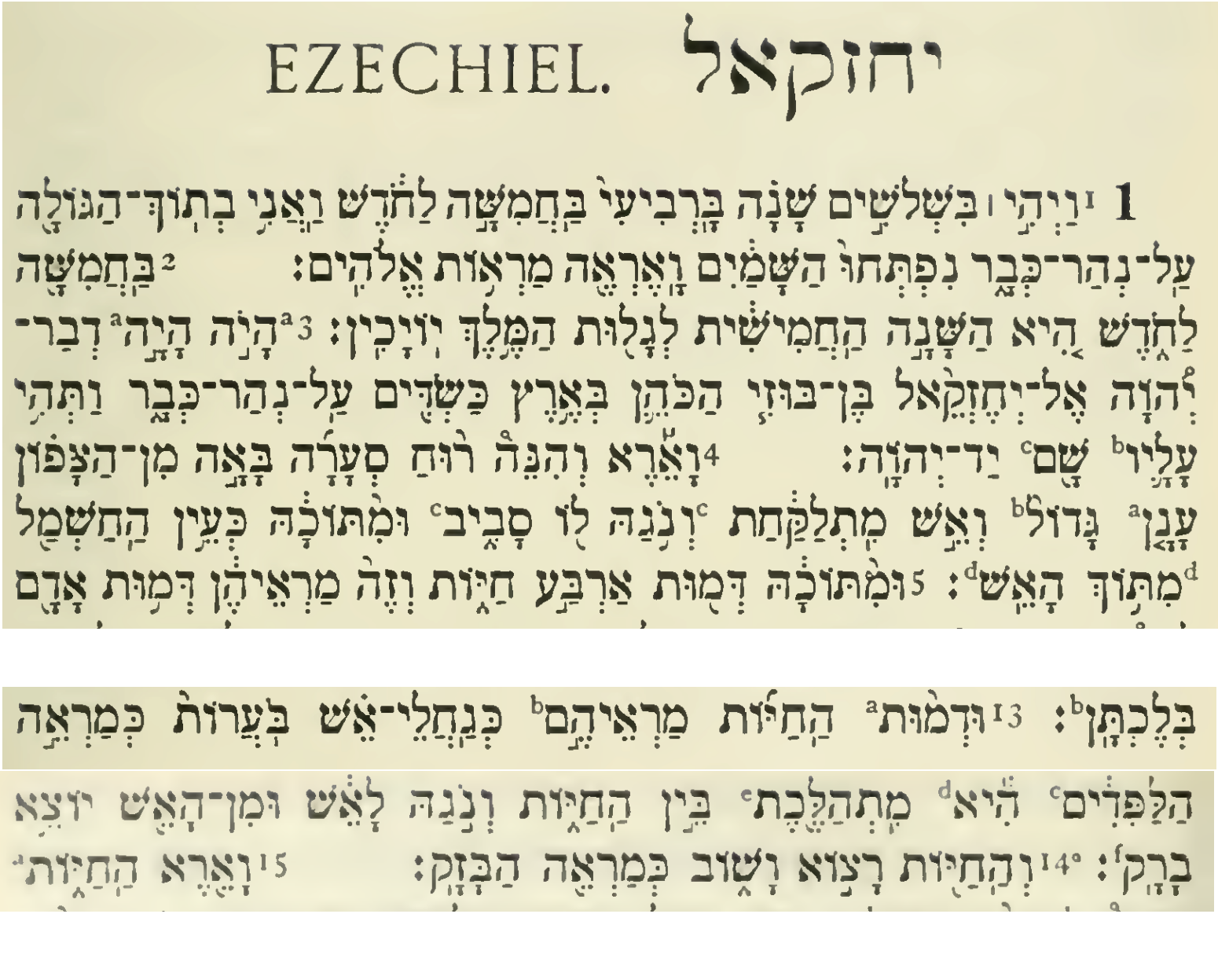

EZECHIEL. יחזקאל

1 וַיְהִי בִּשְׁלֹשִׁים שָׁנָה בָּרְבִיעִי בַּחֲמִשָּׁה לַחֹדֶשׁ וַאֲנִי בְתוֹךְ־הַגּוֹלָה
עַל־נְהַר־כְּבָר נִפְתְּחוּ הַשָּׁמַיִם וָאֶרְאֶה מַרְאוֹת אֱלֹהִים׃ 2 בַּחֲמִשָּׁה
לַחֹדֶשׁ הִיא הַשָּׁנָה הַחֲמִישִׁית לְגָלוּת הַמֶּלֶךְ יוֹיָכִין׃ 3 הָיֹה הָיָה דְבַר־
יְהוָה אֶל־יְחֶזְקֵאל בֶּן־בּוּזִי הַכֹּהֵן בְּאֶרֶץ כַּשְׂדִּים עַל־נְהַר־כְּבָר וַתְּהִי
עָלָיו שָׁם יַד־יְהוָה׃ 4 וָאֵרֶא וְהִנֵּה רוּחַ סְעָרָה בָּאָה מִן־הַצָּפוֹן
עָנָן גָּדוֹל וְאֵשׁ מִתְלַקַּחַת וְנֹגַהּ לוֹ סָבִיב וּמִתּוֹכָהּ כְּעֵין הַחַשְׁמַל
מִתּוֹךְ הָאֵשׁ׃ 5 וּמִתּוֹכָהּ דְּמוּת אַרְבַּע חַיּוֹת וְזֶה מַרְאֵיהֶן דְּמוּת אָדָם

בְּלֶכְתָּן׃ 13 וּדְמוּת הַחַיּוֹת מַרְאֵיהֶם כְּגַחֲלֵי־אֵשׁ בֹּעֲרוֹת כְּמַרְאֵה
הַלַּפִּדִים הִיא מִתְהַלֶּכֶת בֵּין הַחַיּוֹת וְנֹגַהּ לָאֵשׁ וּמִן־הָאֵשׁ יוֹצֵא
בָרָק׃ 14 וְהַחַיּוֹת רָצוֹא וָשׁוֹב כְּמַרְאֵה הַבָּזָק׃ 15 וָאֵרֶא הַחַיּוֹת

Supplementary Figure S2: Ezekiel's Vision in the critical edition of *Biblia Hebraica*

(Kittel, 1906, pp. 744-745).

## Appendix 2: Transliteration and Translation of Assyrian Astrological Reports for the Candidate Auroral Records

We examined the available original cuneiform tablets in the British Museum for each of the three candidate auroral records and created trace copies as the basis for transliterations and translations. Because tablet 80-7-19,19 for the report of a "red sky" could not be delivered when we requested an examination, we made transcriptions and translations, relying upon existing copies by Pinches (1882, p. 10) and Harper (1896, p. 439), as well as the transliteration of Hunger (1992, p. 173).

As shown in the Section 2, two accounts show that the astrological reports are based on the regular observations and drafted with astrological interpretations upon the occurrence of cited celestial events. Unless otherwise endorsed, it is surmised that quoted protases in the reports indicate observed celestial events.

Tablet R1: Rm211 (Preserved in the British Museum, RMA 275; Hunger 1992, no. 35). Copy of 1-6 and r.1 (Figure 1a)

Tablet R1 (Rm211; Figure 1a) is one of the astrological reports of the scholar Issār-šumu-ēreš to Assyrian kings Esarhaddon and Assurbanipal (collected by Hunger 1992, pp. 4-21). The following transliteration and restoration are based on the photos of the tablets © Hisashi Hayakawa (taken courtesy of the Trustees of the British Museum) and the transliteration and restoration of Hunger (1992, p. 19). Hunger's restoration, except for the name of the sender, was proposed by Neugebauer and Weidner (1915, pp. 56-61) on the basis of three astrological omens in the tablet VAT 9417. Its text was edited there, but the tablet number was later shown by Weidner (1941, p. 175).

Transliteration

1 1 *ina* $^{iti}$⸢SIG$_4$⸣ *a-k*[*u-ku-tum* MÚ-*uḫ*]

2 MUNUS.KÚR *ina* [KUR GÁL-*ši*]

3 1 *a-ku-*⸢*ku-*⸣*tum* I[M.U$_{18}$.LU *rak-bat*]

4 UD-*mu* [ŠÚ-*am*]

5 1 *a-ku-ku-tum ina* [AN.PA *it-ta-na-an-paḫ*]

6 KUR *i*[*n-neš-ši* KI.MIN TUR-*ár*]

r.1 *ša* LÚ[.GAL–A.BA]

Translation

1 If in the month Sivan (III) a red gl[ow blazes,]
2 [there will be] hostility in [the land.]
3 If a red glow [rides south wind,]
4 the day[2] will [grow cloudy.]
5 If a red glow [keeps blazing] at [the zenith,]
6 the land will [fall into anarchy, or the ditto will be made smaller.]
r.1 From the [Chief Scribe]

Tablet R2: K748 (Preserved in the British Museum, RMA 248; Hunger 1992, no. 78). Copy of 1-2 (Figure 1b)

The transliteration is based on the photos of the tablets © Hisashi Hayakawa (taken courtesy of the Trustees of the British Museum) and the transliteration of Hunger (1992, p. 45). The sign group *sa-a-mu* in l.1 is an erroneous Akkadian syllabic script for the Sumerogram SA$_5$. It should be read *sāmtu* because *erpetu* (IM.DIRI) is not a masculine noun but rather a feminine noun. The sign groups *it-ta-na-áš-kan* and *šá-a-ru* on l.2 are Akkadian syllabic script for GAR.GAR-*nu* and TU$_{15}$.

Transliteration

1 1 IM.DIRI | SA$_5$ *ina* AN-*e*
*sa-a-mu*
2 GAR.GAR-*nu* TU$_{15}$ ZI-*a*
*it-ta-na-áš-kan* *šá-a-ru*
blank
r.1 *ša* $^{md}$PA—PAB.MEŠ—SU

Translation

1 If a red cloud keeps being placed in the sky,
2 a wind will rise.

r.1 From Nabû-aḫḫē-erība.

[2] Akkadian *ūmu* (UD-*mu*) means not only daytime but also day as a unit of time (from one sunset to the next sunset) (Roth, 2010, pp. 138-155). Therefore, we cannot narrow down its time range to just daytime (daylight).

Tablet R3: 80-7-19,19 (Preserved in the British Museum, RMA 267A; Hunger 1992, no. 309)

The transliteration is based on copies (Pinches 1882, p. 10; Harper 1896, p. 439) and the transliteration of Hunger (1992, p. 173).

Transliteration

1 1 AN-*e* SA$_5$ *ma-ḫi-iṣ* ḪÉ.GÁL *ina* KUR GÁL-*ši*

(The transliteration of 2-r.7 is omitted)

r.8 *šá* $^{m}$*za-kir*

Translation

1 If red covers the sky, there will be abundance in the land.

(The translation of 2-r.7 is omitted)

r.8 From Zākiru

**Appendix 3: Comparable Auroral Reports during the Known Space Weather Events**

As pre-modern auroral reports are frequently vague and subject to critical analysis (Silverman, 2006; Usoskin *et al*., 2017), it is always important to reference parallel auroral reports during known space weather events (Stephenson *et al*., 2019).

In tablet R1 (tablet Rm211), the "red glow (*akukūtu*)" was reportedly blazing at the zenith and moved from south to north, as associated with a south wind. Likewise, in tablet R3 (tablet 80-7-19,19), it was reported: "red covers the sky (AN-*e* SA$_5$ *ma-hi-iṣ*)". Although aurorae in mid to low latitudes are frequently expected along the equatorward horizon, aurorae during extreme space weather events can reportedly cover the entire sky and even show poleward retreat.

On 1872 Feb. 4, an extreme space weather event caused global auroral displays visible down to ≈ 19° or even more equatorward (Silverman, 2008). Hall (1872) reported an auroral display at Barnstaple (N51°05′, W004°04′; 55.0° MLAT) as follows: "Up to 6.55, no auroral light was distinguishable in the northern portion of the sky; but at that moment there was a sudden outburst of rays from the central point, covering the entire heaven in every quarter, several of the ray's in the E. and E.N.E. being, however, especially remarkable for their width and colour". Here, an aurora was first seen in the southern portion of the sky and then covered the entire sky. This report is similar to the auroral report at London (N51°37′, W00°08′, 55.0° MLAT): "in the southwest there was an intense glare of red covering a very large extent" (Loomis, 1860, p.389). These descriptions are comparable with those in R1 and R3.

Tablet R2 (K748) reported the "red cloud (*erpetu* (IM.DIRI) *sāmtu* ($SA_5$))" being placed in the sky. Aurorae were frequently described as a "red cloud" even in early-modern observations. Reporting the space weather event on 1859 Aug. 28, which precedes the famous Carrington storm (Tsurutani *et al.*, 2003; Cliver and Dietrich, 2013), William Dawson in Henry Co., Indiana (N40°00′ W85°15′, 50.6° MLAT), states as follows: "Aug, 28th, about 9 P. M. a red cloud covered a large portion of the eastern sky with a similar one in the N. W." (Loomis, 1860, p. 345). This report is also consistent with the description of a red cloud in R2.

Considering their relatively long duration, stable auroral red (SAR) arcs are plausible in this case, as they are broad in shape and last for more than a few hours without significant motion (Kozyra *et al.*, 1997). A red-dominated aurora is also a possible candidate (Miyaoka *et al.*, 1990). Otherwise, the report may refer to volcanic aerosols, although Mesopotamia is far from active volcanoes.

Tablet R3 (tablet 80-7-19,19) reported 'red covers the sky'. Although this text is very short, the coverage of redness in the sky is consistent with auroral displays covering a large part of the sky, either by SAR arcs or normal red aurorae (*c.f.*, Kozyra *et al.*, 1997; Miyaoka *et al.*, 1990).

These parallel reports demonstrate that the candidate auroral observations in the Assyrian Astrological Reports are consistent with early modern visual auroral observations during known space weather events. Given that the MLATs of Assyria and Babylonia were ≈ 8° more poleward than those in modern times (see Figure 3), it is plausible that the auroral displays were seen more frequently and more dramatically in these regions in the early 7th century BCE.